# A Summary for Lunchtime

The 'quantum counterfactuality' is one of the most striking counterintuitive effects predicted by quantum mechanics. This paper shows that the counterfactual effect is not merely an interesting academic theme, but that it can also provide practical benefits in everyday life. Based on the quantum counterfactual effect, the task of a secret key distribution between two remote parties can be accomplished even when no particle carrying secret information is in fact transmitted. The secret key obtained in this way may be used for secure communications such as internet banking and military communications. This paper also shows that, in some cases, the mere possibility that an eavesdropper can commit a crime is sufficient to detect the eavesdropper, even though the crime is not in fact carried out. (It is still unclear whether the mere possibility that the eavesdropper can commit a crime is sufficient to punish him/her, even though he or she does not in fact do it!) In a sense, part of the story of the SF film *Minority Report* seems plausible.

# Counterfactual Quantum Cryptography

Tae-Gon Noh

*Electronics and Telecommunications Research Institute, Daejeon 305-700,*

*Republic of Korea*

(Version 2; Dated: September 26, 2008)

**Quantum cryptography allows one to distribute a secret key between two remote parties using the fundamental principles of quantum mechanics. The well-known established paradigm for the quantum key distribution relies on the actual transmission of signal particle through a quantum channel. This paper shows that the task of a secret key distribution can be accomplished even though a particle carrying secret information is not in fact transmitted through the quantum channel. The proposed protocols can be implemented with current technologies and provide practical security advantages by eliminating the possibility that an eavesdropper can directly access the entire quantum system of each signal particle.**

According to quantum mechanics, events that might have occurred can have actual physical effects, even though they do not in fact occur *(1)*. What has been termed as an interaction-free measurement *(2–4)* is a typical example of such striking counterfactual phenomena: the presence of an object can be determined without a photon being scattered by the object. It has also been shown that the outcome of a quantum computation can sometimes be inferred without the running of a computer *(5–8)*. This counterfactual computation exhibits a surprising counterintuitive quantum computational effect, but it seems that it does not have a practical advantage for a specific computational purpose in its present form. Here we apply the fundamental concept of quantum counterfactuality to a real-world communication task. We present a novel class of counterfactual protocols of quantum cryptography *(9–12)* that relies on the 'non-transmission' of a signal particle (the carrier of secret information): the mere possibility for signal particles to be transmitted is sufficient to create a secret key.

Quantum cryptography, also known as quantum key distribution (QKD), is considered to be a method of providing unconditional security in communications between two remote parties ('Alice' and 'Bob' in the example below). It allows, in principle, the seamless distribution of a secret key that can be used efficiently as a one-time pad. Any attempt by an eavesdropper ('Eve') to gain information about the key can be not only protected against, but also discovered based on the laws of quantum mechanics.

The previous protocols of QKD require the transmission of signal particles through a quantum channel. For instance, Alice prepares a single photon in a quantum state and sends it to Bob. Bob performs a measurement on the received signal photon. Alice and Bob then obtain a perfectly correlated secret key by carrying out the subsequent classical procedures of basis reconciliation, error correction, and privacy amplification. Entanglement-based protocols *(13–14)* also require the transmission of signal particles. To date, all of the proposed and demonstrated QKD protocols of which we are aware fall into the paradigm of 'signal particle transmission'. (All communication methods, either classical or quantum, proposed thus far may fall into this paradigm.)

We present a different approach based on the quantum counterfactual effect. Figure 1 shows the typical architecture of the proposed QKD system. The protocol is initiated by triggering the single-photon source S, which emits a short optical pulse containing a single photon. The single-photon pulse passes through the optical circulator C and is then split by the beam splitter BS. The polarization state of the single-photon pulse is chosen at random to have either horizontal polarization $|H\rangle$ representing the bit value '0', or vertical polarization $|V\rangle$ representing '1'. According to the chosen bit value, the initial quantum state after the BS is given by one of the two orthogonal states

$$|\phi_0\rangle = \sqrt{T}|0\rangle_a|H\rangle_b + i\sqrt{R}|H\rangle_a|0\rangle_b \qquad (1)$$

$$|\phi_1\rangle = \sqrt{T}|0\rangle_a|V\rangle_b + i\sqrt{R}|V\rangle_a|0\rangle_b \qquad (2)$$

where $a$ and $b$ represent, respectively, the path toward Alice's Faraday mirror (FM) and the path toward Bob's site, and where $|0\rangle_k$ denotes the vacuum state in the mode $k = a, \ b$. $R$ and $T = 1 - R$ are the reflectivity and transmissivity of the BS, respectively.

Bob also randomly chooses one of the two polarizations representing his bit value. Bob blocks the optical path $b$ of the single-photon pulse if the polarization of the pulse is identical to his polarization. The blocking of optical path $b$ in such a polarization-selective way can be suitably accomplished, for instance, using the setup depicted in Bob's site (Fig. 1). If an optical pulse incident on Bob's site is horizontally polarized, it passes through the polarizing beam splitter PBS and goes directly to the high-speed optical switch SW. However, if the pulse is vertically polarized, it is first reflected by the PBS, passes through the optical loop OL, and then goes to the SW. Therefore, through accurate control of the switch timing, Bob can effectively switch the polarization state to the detector D3.

On the other hand, if the single-photon pulse has a polarization orthogonal to Bob's, its optical path $b$ is not affected by the SW. Hence, a split pulse travelling through path $b$ may be reflected by the FM in Bob's site and is returned back to the BS. Here, when a split pulse is returned back to the BS, the total optical path length along path $b$ is identical for the two orthogonal polarization states although the two states experience different paths in Bob's site. The function of the two FMs is to transform the polarization state into its orthogonal, to offset possible birefringence effects automatically in the optical paths of the interferometer. It is also assumed that the detectors shown in Fig. 1 can measure the polarization state of a detected photon *(15)*.

The interferometer can be stabilized using feedback control; therefore, if Alice's and Bob's bit values differ, the photon leaves the interferometer going toward detector D2 with certainty owing to the interference effect (the phase difference is $\pi$ radians between the two paths $a$ and $b$). If, however, Alice's and Bob's bit values are equal, the split pulse in path $b$ is blocked by detector D3 and the interference is destroyed. In this case, there are three possibilities for a single photon; (i) the photon travels through path $a$ and is detected at detector D1 with probability $RT$ ; (ii) the photon travels through path $a$ and is detected at detector D2 with probability $R^2$; (iii) the photon goes to Bob through path $b$ and is detected at detector D3 with probability $T$. After the detection of a photon is completed, Alice and Bob tell each other whether or not each of the detectors clicked. If D2 or D3 clicks, they also announce both the detected polarization state and the initial polarization states that were chosen. This is intended to detect Eve's intervention by monitoring the correct operation of the interferometer. Additionally, if D1 clicks alone, Alice compares the detected polarization state to her initial polarization state: if they are consistent, she does not reveal any information about the polarization states; otherwise, she also announces her measurement results.

Alice and Bob can then establish an identical bit string (a 'sifted key') by selecting only the events for which D1 alone detects a photon with a correct final polarization state. They disregard all other events, including events in which multiple detectors click or where no detector clicks (those events can be monitored to improve the security). The overall efficiency of creating a sifted key bit is $RT/2$. As Alice announces only the fact that a photon was detected at D1 with a correct polarization state, the bit information is not revealed to Eve. As in conventional QKD protocols, Alice and Bob can estimate an error rate using small portions of the sifted key obtained this way in order to detect Eve's intervention. A shorter key remaining after the error estimation step may undergo error correction and privacy amplification to become the secure final key.

In the discussion above, a sifted key is created by post-selecting only the events during which a single photon is detected at D1. Thus, in ideal cases, the photons used to create a sifted key have not travelled through path $b$ but only through path $a$ (if the photons have travelled through the path *b*, they must have been detected at D3). The task of a secret key distribution, therefore, can be accomplished without any photon carrying secret information being sent through the quantum channel (path *b*). A photon that carries secret information has been confined from its birth to death within Alice's secure station, and Eve can never access the photon. Formally speaking, when Alice's and Bob's bit values are equal, the initial state $|\phi_0\rangle$ collapses to one of the two states, $|0\rangle_a |H\rangle_b$ or $|H\rangle_a |0\rangle_b$, and the initial state $|\phi_1\rangle$ collapses to $|0\rangle_a |V\rangle_b$ or $|V\rangle_a |0\rangle_b$, due to Bob's measurement. To create a sifted key bit, Alice and Bob use only two states, $|H\rangle_a |0\rangle_b$ and $|V\rangle_a |0\rangle_b$, among the four collapsed states. Hence, Bob in fact extracts a secret key from the non-detection events.

The security of the proposed protocol can be understood by a no-cloning principle *(16)* of orthogonal states in a composite system which consists of two subsystems. It is known that there are cases in which orthogonal states cannot be cloned if the subsystems are only available one after the other *(17–19)*. However, an important point of the present protocol is that Eve can only access one subsystem (path *b*) while she can never access the other subsystem (the path *a*). Hence, we present here another type of no-cloning principle for orthogonal states: if reduced density matrices of an available subsystem are non-orthogonal and if the other subsystem is not allowed access, it is impossible to distinguish two orthogonal quantum states without disturbing them. Let $|\Psi_0\rangle$ and $|\Psi_1\rangle$ be two normalized pure states of a quantum system *AB* composed of two subsystems, *A* and *B*. According to the Schmidt decomposition,

$$|\Psi_0\rangle = \sum_i \lambda_i |i_A\rangle |i_B\rangle \qquad (3)$$

$$|\Psi_1\rangle = \sum_j \lambda_j |j_A\rangle |j_B\rangle \tag{4}$$

where $|i_A\rangle$ ($|i_B\rangle$) and $|j_A\rangle$ ($|j_B\rangle$) are orthonormal states for the subsystem $A$ ($B$), and where $\lambda_i$ and $\lambda_j$ are the Schmidt coefficients. We also suppose that a unitary operator $U$ acts only on the product space of the subsystem $B$ and Eve's measuring apparatus that is in an initial normalized state $|m\rangle$. To conceal Eve's intervention, the states $|\Psi_0\rangle$ and $|\Psi_1\rangle$ should be left undisturbed after the unitary evolution:

$$U(|\Psi_0\rangle|m\rangle) = |\Psi_0\rangle|m_0\rangle$$
$$U(|\Psi_1\rangle|m\rangle) = |\Psi_1\rangle|m_1\rangle \tag{5}$$

Here, $|m_0\rangle$ and $|m_1\rangle$ are the final states of Eve's measuring apparatus. As $U$ does not act on subsystem $A$, equation (5) becomes

$$U(|i_B\rangle|m\rangle) = |i_B\rangle|m_0\rangle$$
$$U(|j_B\rangle|m\rangle) = |j_B\rangle|m_1\rangle \tag{6}$$

Thus, by unitarity

$$\langle i_B | j_B \rangle = \langle i_B | j_B \rangle \langle m_0 | m_1 \rangle \tag{7}$$

from which it follows that either $|m_0\rangle = |m_1\rangle$, or $\langle i_B | j_B \rangle = 0$ for all $i$ and $j$. The condition $\langle i_B | j_B \rangle = 0$ for all $i$ and $j$ implies that reduced density matrices of the subsystem $B$, $\rho_s(B) = \mathrm{Tr}_A[|\Psi_s\rangle\langle\Psi_s|]$, are orthogonal ($\mathrm{Tr}[\rho_0(B)\rho_1(B)] = 0$). Therefore, provided that the reduced density matrices of the available subsystem $B$ are non-orthogonal, Eve cannot gain any information without disturbing the states $|\Psi_0\rangle$ or $|\Psi_1\rangle$, even when the states are orthogonal. It can be verified from Eq. 1 and 2 that the reduced density matrices of the

available subsystem (the path $b$) are non-orthogonal. That is,

$\mathrm{Tr}[\rho_0(\mathrm{path}\,b)\rho_1(\mathrm{path}\,b)] = R^2 \neq 0$ , where $\rho_s(\mathrm{path}\,b) = \mathrm{Tr}_{\mathrm{path}\,a}[|\phi_s\rangle\langle\phi_s|]$ . For $R = 0$ ,

however, the states $|\phi_0\rangle$ and $|\phi_1\rangle$ can be distinguished without disturbance, which is consistent with our intuition.

In conventional QKD protocols relying on the transmission of a signal particle, Eve can fully access, individually or coherently, signal particles sent through the quantum channel. In the present protocol, however, Eve cannot access the entire quantum system of each signal particle, but only part of the quantum system *(20)*. This distinctive property naturally leads to practical security advantages in various situations (see Appendix).

A complete analysis of the QKD security, including various experimental

imperfections, is left for future study *(21–25)*. However, it is worthwhile to point out here

that the proposed protocol provides clear security advantages for cases in which weak

coherent pulses with nonzero multiphoton probabilities are used for practical

implementation in place of single-photon pulses. First, Eve cannot determine the number of

photons in each pulse because she is not allowed to access path $a$. Furthermore, it is

impossible for Eve to measure even the number of photons travelling through the quantum

channel (path $b$), provided that she does not disturb the states. Eve obtains 'which-path'

information through the photon number measurement in path $b$, and she destroys the interference. Hence, Eve may cause detection errors, and she may be detected due to the photon number measurement itself. Thus, the proposed protocol is inherently robust against the so-called 'photon number splitting' attack *(26, 27)*. Second, Eve cannot split a photon when all of the photons in the pulse travel through path $a$. That is, if all of the photons are detected at D1 after travelling through path *a*, the bit information is not revealed to Eve, even when a multiphoton pulse is used. Finally, Eve cannot obtain a copy of the quantum state even when she succeeds in splitting a photon: she remains limited by the no-cloning theorem.

Every component and device needed for the proposed QKD system is currently available. We hopefully expect experimental demonstration of the proposed protocol will appear soon. Additionally, we have considered a Michelson-type interferometer using two orthogonal states merely because it is simple and feasible for practical applications. It is clear that the present protocol can be modified to incorporate the features of other QKD protocols. For instance, instead of using two orthogonal polarization states, it is possible to use either four polarization states as in the BB84 protocol *(10)* or two non-orthogonal states as in the B92 protocol *(11)*. These protocols can also be correctly implemented using a Mach-Zehnder type interferometer. Perhaps various protocols and implementation schemes will appear within the counterfactual paradigm, where the central concept is the non-transmission of a signal particle.

This work was partially supported by the IT R&D program of MKE/IITA (2005-Y-001-04 and 2008-F-035-01).

# References and Notes

1. R. Penrose, *Shadows of the Mind* (Oxford Univ. Press, New York, 1994).

2. A. C. Elitzur, L. Vaidman, *Found. Phys.* **23**, 987 (1993).

3. P. G. Kwiat *et al.*, *Phys. Rev. Lett.* **83**, 4725 (1999).

4. T.-G. Noh, C. K. Hong, *Quantum Semiclass. Opt.* **10**, 637 (1998).

5. R. Jozsa, in *Lecture Notes in Computer Science,* C. P. Williams, Ed. (Springer-Verlag, Berlin, 1999), vol. 1509, pp. 103–112.

6. G. Mitchison, R. Jozsa, *Proc. R. Soc. Lond. A* **457**, 1175 (2001).

7. O. Hosten, M. T. Rakher, J. T. Barreiro, N. A. Peters, P. G. Kwiat, *Nature* **439**, 949 (2006).

8. L. Vaidman, *Phys. Rev. Lett.* **98**, 160403 (2007).

9. S. Wiesner, *SIGACT News* **15**, 78 (1983).

10. C. H. Bennett, G. Brassard, in *Proceedings of the IEEE International Conference on Computers, Systems, and Signal Processing, Bangalore, India* (IEEE press, New York, 1984), pp. 175–179.

11. C. H. Bennett, *Phys. Rev. Lett.* **68**, 3121 (1992).

12. N. Gisin, G. Ribordy, W. Tittel, H. Zbinden, *Rev. Mod. Phys.* **74**, 145 (2002).

13. A. K. Ekert, *Phys. Rev. Lett.* **67**, 661 (1991).

14. C. H. Bennett, G. Brassard, N. Mermin, *Phys. Rev. Lett.* **68**, 557 (1992).

15. This can be conducted simply by ensuring that each of the detectors has a polarizing beam splitter and two conventional single-photon detectors.

16. W. K. Wootters, W. H. Zurek, *Nature* **299**, 802 (1982).

17. T. Mor, *Phys. Rev. Lett.* **80**, 3137 (1998).

18. M. Koashi, N. Imoto, *Phys. Rev. Lett.* **79**, 2383 (1997).

19. L. Goldenberg, L. Vaidman, *Phys. Rev. Lett.* **75**, 1239 (1995).

20. Because of this property, perhaps the present protocol will be proved to be robust against coherent attacks *(12)*.

21. D. Mayers, in *Lecture Notes in Computer Science*, N. Koblitz, Ed. (Springer-Verlag, Berlin, 1996), vol. 1109, pp. 343–357.

22. H.-K. Lo, H. F. Chau, *Science* **283**, 2050 (1999).

23. P. W. Shor, J. Preskill, *Phys. Rev. Lett.* **85**, 441 (2000).

24. H. Inamori, N. Lütkenhaus, D. Mayers, *Eur. Phys. J. D* **41**, 599 (2007).

25. D. Gottesman, H.-K. Lo, N. Lütkenhaus, J. Preskill, *Quant. Info. Comp.* **4**, 325 (2004).

26. N. Lütkenhaus, *Phys. Rev. A* **61**, 052304 (2000).

27. G. Brassard, N. Lütkenhaus, T. Mor, B. C. Sanders, *Phys. Rev. Lett.* **85**, 1330 (2000).

# Figures

**Fig. 1.** Schematic of the proposed QKD system. A single-photon pulse entering a Michelson-type interferometer is split into two pulses by a beam splitter BS and travels through two paths *a* and *b*. The interferometer is adjusted using an optical delay OD. Therefore, if the bit values chosen at random by Alice and Bob are different, the two split pulses are recombined in the BS and the single photon is detected at detector D2 with certainty as a result of constructive interference. However, if the two bit values are equal, a split pulse going through path *b* is blocked by detector D3. Consequently, the interference is destroyed and the photon can be detected at detector D1 with a finite probability. In this case, the photon has been completely isolated from the outside of Alice's secure station, as it has traveled through only path *a*.

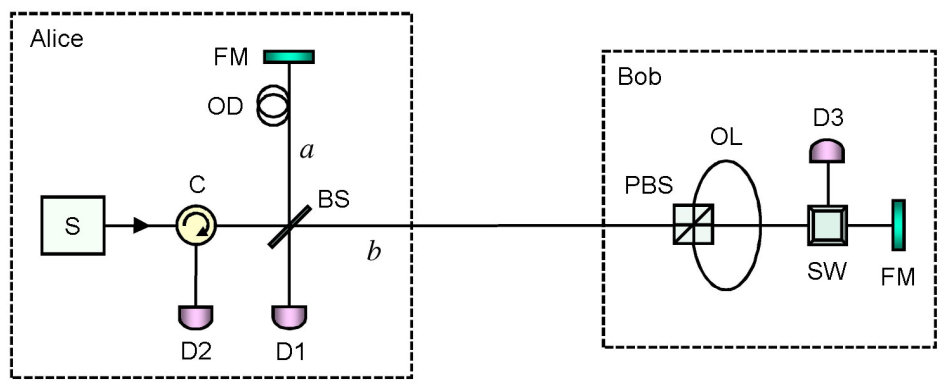

# Appendix

In terms of security against a simple intercept-resend (I-R) attack, when Eve uses this strategy in conventional QKD protocols relying on the transmission of a signal particle, she can always intercept the signal particle, perform a measurement as Bob does, and then resend to Bob a fake particle compatible with her measurement result. However, in the present protocol, Eve sometimes fails to intercept a photon, as it can travel through path $a$, to which Eve does not have access. We assume that Eve, in this case, would not send a photon, which is equivalent to sending a vacuum state. We also suppose that Eve randomly measures one polarization component as Bob does. Hence, if an optical pulse has a polarization state orthogonal to Eve's, it would simply pass through Eve's apparatus without disturbance. Meanwhile, when Eve is able to detect a photon, she would send a fake photon with polarization identical to the detected polarization in order to minimize changes of the detection probabilities at D1, D2, and D3.

Using this simple I-R strategy, Eve may obtain the following results. First, if Alice's and Bob's polarizations are equal, the detection probabilities at D1, D2, and D3 do not change as if Eve is absent. This is true regardless of Eve's polarization choice. Hence, Eve's intervention is not noticed in this case. However, Eve also obtains no information about the bit value, even when the photon is detected at D1.

Next, if Alice's and Bob's polarizations are orthogonal, there are two possibilities: (I) Eve's polarization is orthogonal to Alice's; (II) Eve's polarization is identical to Alice's. In case (I), the interference is preserved and the photon is detected at D2 with certainty. Alice and Bob discard these events. Eve's intervention is not noticed, and Eve obtains no information. In case (II), however, the interference is destroyed and the detection probabilities change significantly compared to those cases in which Eve is absent: (i) If the photon travels through path $a$ and is detected at D1, Alice and Bob experience a bit error (this happens with probability $RT$). Eve obtains no information about the bit value. (ii) If the photon is detected at D2 through path $a$, Alice and Bob discard this event. Eve's intervention is not noticed and she obtains no information. (iii) If the photon travels through path $b$ with probability $T$, the photon is detected by Eve (Eve knows her polarization is equal to Alice's). Eve would send to Bob a fake photon with a compatible polarization state. The photon is then returned to Eve (Eve may detect the photon again and knows at this stage that her polarization is orthogonal to Bob's). Eve would resend a photon toward the BS in Alice's site (otherwise, Eve must be detected in ideal cases due to the photon loss): if the photon is finally detected at D1, Alice and Bob experience a bit error (this occurs with probability $TR$); if the photon is finally detected at D2, Alice and Bob discard this event (Eve's intervention is not noticed and she obtains no information).

Overall, in this type of I-R attack, the probability that the photon is detected at D1 doubles, resulting in $RT$; however, half of the events are errors, i.e., a 50% error rate exists in the sifted key. Hence, the sifted key is completely corrupted by the attack. In addition, Eve has no chance to learn about the bit value, i.e., Eve's information is 0%. In order to gain some information, Eve can modify the strategy at the expense of safety: she may transform the polarization state into its orthogonal before she sends a photon. By doing this, the probability of creating a sifted key bit also doubles resulting in $RT$; however, this time a 25% error rate exists in the sifted key and 25% of Eve's information. However, with this modification, Eve may cause additional detection errors at D3. That is, a fake photon having the transformed polarization state will be detected at D3, even when Alice's and Bob's polarizations are orthogonal. The overall probability of this error occurring is $T/4$. Thus, Eve's intervention is more easily detected in this modified I-R attack.

These results can be compared with the BB84 protocol in which, if Eve uses an I-R strategy, the error rate in a sifted key is 25% and Eve's information is 50%. Thus, the proposed protocol is more robust against an I-R attack. Additionally, Alice and Bob can detect Eve's intervention more easily by monitoring only the sifted key creation rate without the additional effort of calculating the error rate.

Ideally, a sifted key bit is created only when a signal photon remains within Alice's secure station: the photon travels through path $a$ and is finally detected at D1. In contrast to conventional protocols, Eve cannot modify this process using an I-R attack because she cannot intercept the photon. The effect of an I-R attack is simply to add an error bit or a phony bit. This is why the sifted key rate doubles in the presence of an I-R attack.

We now discuss another security advantage of the proposed protocol. To proceed, we define the quantum channel identification (QCI) problem: For a quantum network in which several quantum channels separated in space-time and/or in terms of their wavelength are available, it is supposed that Alice and Bob use only one of them for the key distribution without revealing it publicly. Then, how can Eve determine the correct quantum channel without being noticed? The QCI is of practical importance, as it is a necessary precondition of any eavesdropping attack. For instance, when Eve uses Trojan-horse attacks *(22)* in which she sends light pulses through the quantum channel and analyzes the backscattered light to probe Alice and/or Bob's apparatuses, if Eve does not identify the correct quantum channel before she applies these attacks, she may easily be detected with high probability by means of auxiliary detectors that monitor any light coming through dummy decoy channels.

The QCI problem, however, can be resolved easily in conventional QKD protocols that rely on the actual transmission of a signal particle: as the quantum states of the particle transmission are orthogonal to the vacuum state, Eve can identify the correct quantum channel without disturbing the states. In contrast, in the proposed protocol, if Eve attempts to identify the correct quantum channel by probing the particle transmission, she may cause a bit error with a nonzero probability. Furthermore, this protocol provides the possibility of hiding the quantum channel itself. These features are explained below. First, a case is considered in which Alice's and Bob's polarizations are equal. In this case, the interference is destroyed as optical path $b$ is blocked by Bob. This is indistinguishable from Eve's action. Considering that Eve obtains 'which-path' information by monitoring the particle transmission, she destroys the interference even when she does not disturb the internal state of the photon (e.g., the polarization state). Hence, legitimate users cannot detect Eve's presence. Meanwhile, when the photon transmits the BS with probability $T$ on the first encounter, Eve may succeed in the QCI. When the photon is reflected by the BS with probability $R$ on the first encounter, Eve fails in the QCI: she cannot distinguish the correct channel from dummy channels, as she finds nothing but the vacuum state.

Next, a case in which Alice's and Bob's polarizations are orthogonal is considered. As mentioned earlier, if the interference is preserved, the photon is always detected at D2. However, as Eve's action destroys the interference, the following results are possible: (i) The photon is reflected by the BS on the first encounter and remains within the Alice's station. In this case, Eve only finds the vacuum state and fails in the QCI. When the photon is also reflected on the second encounter with the BS and is detected at D2 with probability $R^2$, legitimate users cannot detect Eve's action. When the photon transmits the BS on the second encounter and triggers D1 with probability $RT$, a bit error occurs and legitimate users can in principle detect Eve by checking the bit error. (ii) The photon transmits the BS on the first encounter and goes to Bob's station. In this case, Eve succeeds in the QCI by probing the particle transmission. Provided that Eve does not disturb the internal state of the photon, the photon may return to the Alice's station. Thus, when the photon also transmits the BS on the second encounter and triggers D2 with probability $T^2$, Eve is not detected. When the photon is reflected on the second encounter with the BS and detected at D1 with probability $TR$, a bit error occurs. Thus, Eve's action can be detected in principle.

It follows from the above discussion that Eve has four possibilities with a single trial of the QCI in this protocol: (i) Eve succeeds in the QCI and is not detected. This happens with probability $P_1 = T/2 + T^2/2$. (ii) Eve succeeds in the QCI and induces a bit error. This happens with probability $P_2 = TR/2$. (iii) Eve fails in the QCI and is not detected. This happens with probability $P_3 = R/2 + R^2/2$. (iv) Eve fails in the QCI and induces a bit error. This happens with probability $P_4 = RT/2$. The probability $P_1$ is a natural measure of the efficiency of the QCI. Eve can achieve up to $P_1 = 1$ in the limit $T \to 1$. Only in this limiting case, Eve can safely identify the correct quantum channel with certainty, as in conventional QKD protocols. Generally for $T < 1$, however, $P_1$ becomes less than 1. In fact, the efficiency of the QCI can be reduced to $P_1 = 0$ by decreasing the transmissivity $T \to 0$ . In case the sifted key creation probability is maximized ( $R = T = 1/2$ ), the efficiency becomes $P_1 = 3/8$ .

It is interesting to consider in more detail case (iv), in which Eve fails in the QCI because she cannot observe anything except the vacuum. Indeed, in this case, Eve did not 'touch' the photon nor modify its internal state; however, she can still be detected with a nonzero probability by remote legitimate users. This we may call the 'counterfactual detection' of an eavesdropper—the mere possibility that Eve can identify the correct quantum channel is sufficient to detect her, even though she does not in fact do it. Eve may not know that she is detected, as she may think nothing has happened. A similar argument can be applied to I-R attacks. That is, in case (i) of (II) in the discussion above, Eve can be detected although she does not in fact intercept the photon.